\documentclass[paper,twocolumn,showpacs,superscriptaddress,prl]{revtex4}%
\usepackage{amsfonts}
\usepackage{amsmath}
\usepackage{amssymb}
\usepackage{graphicx}%
\begin{document}
\title{The influence of the optical Stark effect on chiral tunneling in graphene}
\author{Jiang-Tao Liu}
\email[Electronic address:]{jtliu@semi.ac.cn}
\affiliation{Department of Physics, Nanchang University, Nanchang
330031, China}%
\author{Fu-Hai Su}
\affiliation{Key Laboratory of Materials Physics, Institute of Solid
State Physics, Chinese Academy of Sciences, Hefei 230031, China}
\author{Hai Wang}
\affiliation{Department of Physics, Capital Normal University,
Beijing 100037, China}%
\author{Xin-Hua Deng}
\affiliation{Department of Physics, Nanchang University, Nanchang 330031, China}%

\date{\today}

\begin{abstract}
The influences of intense coherent laser fields on the transport
properties of a single layer graphene are investigated by solving the
time-dependent Dirac equation numerically. Under an intense laser field, the valence band and conduction band states mix via the optical Stark effect. The chiral symmetry of Dirac electrons is broken and the perfect chiral tunneling is strongly suppressed. These properties might be useful in the fabrication of an optically controlled field-effect transistor.
\end{abstract}
\maketitle


Graphene has attracted much attention due to its
remarkable electronic properties \cite{11KS,12KS,13AH}. The
low-energy quasiparticles, which have linear dispersion and
nontrivial topological structure in their wave function, can be
described by using a Dirac-like equation. This unique band structure
of graphene leads to many important potential applications in
nanoelectronics \cite{16VV,17MI,17ZZ,17VH,17PM,17EP}.

One of the peculiar transport phenomena in graphene is the chiral
tunneling \cite{17MI,16VV,A1WR}. In single layer graphene a
perfect transmission through a potential barrier in the normal
direction is expected. This unique tunneling effect can be explained
by the chirality of the Dirac electrons within each valley, which
prevents backscattering in general. This kind of reflectionless
transmission is independent of the strength of the potential, which
limits the development of graphene-based field-effect transistors
(FET) \cite{17MI}. The perfect transmission can be suppressed effectively
when the chiral symmetry of the Dirac electrons is broken. For
instance, in a magnetic field, a quantized transmission can be
observed in graphene \emph{p-n} Junctions \cite{20DA}. Recently, Elias et. al. proposed that the hydrogenation could convert the semimetal graphene into an insulator material \cite{A3DC}.

The intense optical field can also break the chiral symmetry
of Dirac electrons in graphene, e.g., Fistul and Efetov have shown that when the n-p Junctions in graphene is irradiated by an electromagnetic field in the resonant condition, the quasiparticle transmission is suppressed \cite{21MV}. The optical
field control on carrier transport offers several advantages. Optical fields can control not only the charge carriers but also the spin carriers, especially which can be performed over femtosecond time scale. Another fundamental method of optical control is the optical Stark effect (OSE) \cite{1AM,2SS,3CE,4MC,A2DF}.  The OSE in traditional semiconductors is due to a dynamical coupling of excitonic states by an intense laser field. The OSE have shown many useful applications in optoelectronics and spintronics \cite{5MC,6CE,7SS,9WY,10JT}.

In graphene, the valence band and conduction band
states can also mix strongly via OSE. Thus the
chirality of Dirac electrons will be completely changed, or even
disappear.  Unlike the resonant case
\cite{21MV}, in OSE the coherent excitons are virtual excitons,
which exist only when the optical field is present. Thus the
light-induced shift lasts only for the duration of the pump pulse,
which allows for optical gates that might only exist for
femtoseconds. Furthermore, since there is no real absorption in the
nonresonant case, the absorption of photons is quite small and low
power consumption is expected.

\begin{figure}[b]
\includegraphics[width=0.98\columnwidth,clip]{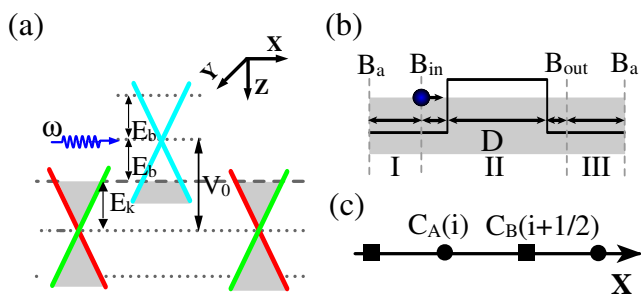}
\caption{(color online). (a) Schematic of the spectrum of Dirac
electrons in single-layer graphene. The optical field is propagated
perpendicular to the layer surface and and is linearly polarized
along the $Y$ direction. (b) Schematic of the scattering of Dirac
electrons by a square potential. $B_{a}$, $B_{in}$, and $B_{out}$
denote the absorbing boundary, incident  boundary, and output
boundary, respectively.
(c) Schematic of the one-dimensional Yee lattice in graphene.}%
\label{fig1}%
\end{figure}

In this Letter, we study the tunneling rate of
Dirac electrons in graphene through a barrier with an intense
electromagnetic field. We consider a rectangular potential barrier
with height $V_{0}$, width $D$ in the  $X$ direction, and infinite
length in the $Y$ direction [see Fig. 1 (a) and Fig. 1 (b)]. The
Fermi level (dashed lines) lies in the valence band in the barrier
region and in the conduction band outside the barrier. The gray
filled areas indicate the occupied states. The optical field is
propagated perpendicular to the layer surface and is linearly
polarized along the $Y$ direction with a detuning $\Delta_{0}=2
E_{b}-\hbar \omega$. We choose $\Delta_{0}>0$ to ensure that there
is no interband absorption inside the barrier. Meanwhile, $\hbar
\omega \ll 2 E_{k}$ is used to guarantee that the influence of the
optical field outside the barrier can be neglected.

Since the Coulomb interaction between electrons and holes in OSE is negligible when the detuning is large \cite{4MC,5MC},
we did not take into account the electron-hole Coulomb interaction or many body effect in our calculation. Thus, neglecting the scattering between different valleys,  the scattering process of Dirac electrons in $K$ point is described by the time-dependent Dirac equation
\begin{equation}
i\hbar \frac{\partial }{\partial t}\mathbf{\Psi }\left(
\mathbf{r},t\right) =\left[ \mathbf{H}_{0}+V_{0}\left(\mathbf{r}
\right) \mathbf{ I}+\mathbf{H}_{int}\right] \mathbf{\Psi
}\left(\mathbf{r},t\right),\label{eq1}
\end{equation}
where $\mathbf{\Psi }\left( \mathbf{r}, t\right)=[C_{A}(\mathbf{r},
t),C_{B}(\mathbf{r},t)] $ is the wave function,
$\mathbf{H}_{0}=-i\hbar v_{F}\mathbf{\sigma}\bullet \nabla$ is the
unperturbed Dirac Hamiltonian,
$\mathbf{\sigma}=(\sigma_{x},\sigma_{y})$ are the Pauli matrices,
$v_{F}\approx 10^{6}m/s$ is the Fermi velocity, $V_{0}(\mathbf{r})$
is the height of the potential barrier, $\mathbf{I}$ is the unit
matrix, and $\mathbf{H}_{int}$ is the interaction Hamiltonian.
$\mathbf{H}_{int}$ can write as \cite{28EJ}
\begin{equation}
\mathbf{H}_{int}=-\hbar ev_{F}\left[ A (x,t)\sigma _{x}+A (y,t)\sigma
_{y}\right] =\hbar \left(
\begin{array}{cc}
0 & V_{12}(t) \\
V_{21}(t) & 0%
\end{array}%
\right),
\end{equation}
where $e$ is the electron charge and $[A (x,t),A (y,t)]=[A_{x}e^{i\omega t}, A_{y}e^{i\omega t}]$ are the vector potentials of the electromagnetic field. When the Dirac electrons is incident on the barrier perdenicularly, we can rewrite  Eq. (\ref{eq1}) as a set of partial differential equations

\begin{eqnarray}
i\partial C_{A}(x,t)/\partial t=&-iv_{F}\partial C_{B}(x,t)/\partial
x +V_{0}C_{A}(x,t) \nonumber\\
&+V_{12}(t)C_{B}(x,t), \label{eqO1}\\
i\partial C_{B}(x,t)/\partial t=&-iv_{F}\partial C_{A}(x,t)/\partial
x+V_{0}C_{B}(x,t)\nonumber\\
&+V_{21}(t)C_{A}(x,t). \label{eqO2}
\end{eqnarray}

Since the tunneling time is sub-picosecond and the potential $V_{12}(t)$ and $V_{21}(t)$ vary as fast as the frequency of incident light beams, this scattering process is strongly time-dependent.  In order to study such a strongly time-dependent scattering process, we employ the finite-difference time-domain (FDTD) method to solve Eq. (\ref{eqO1}) and Eq. (\ref{eqO2}) numerically in the time-domain \cite{26KS}. In the traditional FDTD method, the Maxwell's equations are discretized by using central-difference approximations of the space and time partial derivatives. As a time-domain technique, the FDTD method can demonstrate the propagation of electromagnetic fields through a model in real time. Similar to the discretization of Maxwell's equations in FDTD, we denote a grid point of the space and time as $(i,k)=(i\Delta x,k \Delta t)$ [see Fig. 1(c)], and for the any function of space and time $F(i\Delta x,k \Delta t)=F^{k}(i)$.  the first order in time or space partial differential can be expressed as
\begin{eqnarray}
&\frac{\partial F\left(  x,t\right)  }{\partial x}|_{x=i\Delta x}\approx
\frac{F^{k}\left(  i+1/2\right)  -F^{k}\left(  i-1/2\right)  }{\Delta x},\\
&\frac{\partial F\left(  x,t\right)  }{\partial t}|_{t=k\Delta t}\approx
\frac{F^{k+1/2}\left(  i\right)  -F^{k-1/2}\left(  i\right)  }{\Delta t}.
\end{eqnarray}
Thus the Eq. (\ref{eqO1}) and Eq. (\ref{eqO2}) can be replaced by a finite set of finite differential equations

\begin{subequations}
\begin{align}
C_{A}^{k+1/2}(i)&\left[\frac{1}{\Delta t}-\frac{V_{0}(i)}{2i}\right] =\left[\frac{%
1}{\Delta t}+\frac{V_{0}(i)}{2i}\right]C_{A}^{k-1/2}(i) \nonumber \\%
&-\left[ \frac{ v_{F}}{\Delta x}-\frac{V_{12}^{k}(i+1/2)}{2i}\right]
C_{B}^{k}(i+1/2)\nonumber\\%
&+\left[ \frac{v_{F}}{\Delta x}+\frac{V_{12}^{k}(i-1/2)}{2i}\right]
C_{B}^{k}(i-1/2),\label{fdtda}
\end{align}%
\begin{align}
C_{B}^{k+1}&(i+1/2)\left[ \frac{1}{\Delta t}-\frac{V_{0}(i+1/2)}{2i}\right] =%
\left[ \frac{1}{\Delta t}+\frac{V_{0}(i+1/2)}{2i}\right]\times\nonumber \\& C_{B}^{k}(i+1/2)%
-\left[ \frac{v_{F}}{\Delta x}-\frac{V_{21}^{k+1/2}(i+1)}{2i}\right]
C_{A}^{k+1/2}(i+1)\nonumber \\
&+\left[ \frac{v_{F}}{\Delta x}+\frac{V_{21}^{k+1/2}(i)}{2i}%
\right] C_{A}^{k+1/2}(i),\label{fdtdb}
\end{align}%
\end{subequations}
For computational stability, the space
increment $\Delta x$ and the time increment $\Delta t$  need to
satisfy the relation $\Delta x>v_{F}\Delta t$ \cite{26KS}.
Furthermore, the space increment $\Delta x$  must far smaller than
the wavelength of electrons $\Delta x<\lambda_{e}/8$, and the time
increment $\Delta t$ must be far smaller than the period of the
electromagnetic field $T_{l}$.

At the boundary $B_{a}$, one-dimensional Mur absorbing boundary
conditions are used \cite{29GM}. At the input boundary $B_{in}$, a
Gaussian electronic wave packet is injected
\begin{equation}
C_{A}=C_{B}=\frac{1}{\sqrt{2}}\exp \left[ -\frac{4\pi (t-t_{0})^{2}}{\tau
^{2}}\right],
\end{equation}
where $t_{0}$  and $\tau$  denote the peak position and the pulse
width, respectively.

\begin{figure}[t]
\includegraphics[width=0.9\columnwidth,clip]{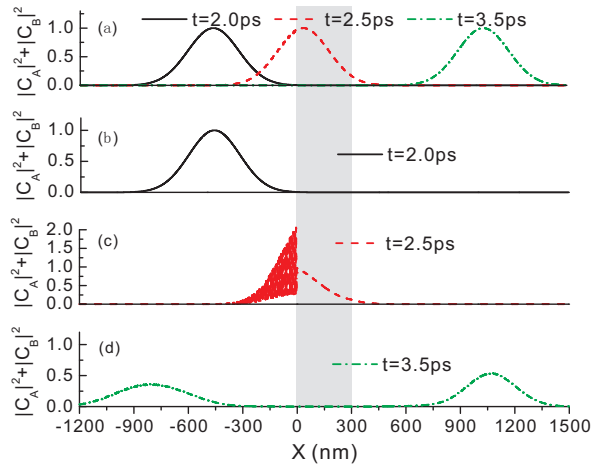}
\caption{(color online). (a) numerical simulations of a wave packet tunneling through a barrier without pump beams. (b)-(d) Time sequence of a wave packet tunneling through a barrier with pump intensity $I_{\omega}=3$
MW/cm$^{2}$, $\Delta_{0}=5meV$, and $D=300$ nm. The light grey shows the barrier area.}
\label{fig2}%
\end{figure}

Thus, by solving Eq. (\ref{fdtda}) and Eq. (\ref{fdtdb}) directly in the time domain we can demonstrate the propagation of a wave packet through a barrier in real time. Numerical simulations are shown in Fig. 2. The following parameters are used in our calculation: the peak position $t_{0}=1.5$ ps, the pulse width $\tau=1.0$ ps, the space increment  $\Delta x=0.1$ nm, the time increment  $\Delta
t=5\times10^{-5}$ ps, and the height of the potential barrier
$V_{0}=400$ meV. When there is no pump beams, a perfect  chiral tunneling can be found [see Fig. 2(a)]. This result is consistent with that of Geim et. al. \cite{17MI}. But when the sample is irradiated by an intense nonresonant laser beam, a reflected wave packet appears [see Fig. 2(d)]. The perfect transmission is suppressed. By analyzing the the transmitted  wave packet and the reflected  wave packet, we can obtain the tunneling rate.

To explain the suppression of chiral tunneling, We first investigate the OSE in the barrier within a rotating-wave approximation \cite{2SS,9WY,10JT}.  Figure 2(a) shows the renormalized band as a
function of momentum $k$ with intensity $I_{\omega}=30$ MW/cm$^{2}$.
In the case of nonresonant excitation, $\hbar \omega <2E_{b}$ and
the dressed states are blue shifted. With increasing detuning, the
light-induced shift decreases, and the dressed states asymptotically
approach the unperturbed states. The intense electromagnetic field
can also induce a strong band mixing. Near the absorption edge, a
maximum fermion distribution function $n_{k}\approx 0.44$ can be
observed [see Fig. 1(b)].

\begin{figure}[t]
\includegraphics[width=0.98\columnwidth,clip]{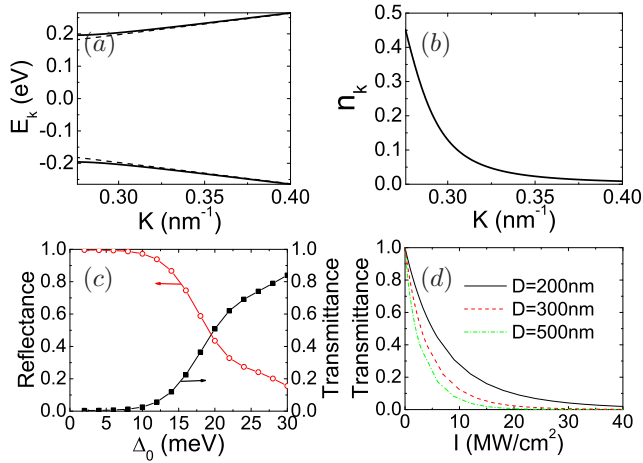}
\caption{(color online). (a) Sketch of the renormalized band
energies (solid lines) and the unperturbed band energies (dashed
lines) as a function of momentum $k$. (b) Sketch of the fermion
distribution function $n_{k}$  as a function of momentum $k$.  (c)
The reflectance (red circles) and the transmittance (black squares)
of the barrier as a function of the detuning for $I_{\omega}=30$
MW/cm$^{2}$ and $D=300$ nm. (d) The transmittance as a function of
pump intensity for $\Delta_{0}=5$ meV with different barrier width.}
\label{fig3}%
\end{figure}

Under intense light beams, the dressed states are strongly mixed
with valence states and conduction states. Therefore, the chiral
symmetry of Dirac electrons in graphene can be broken. For instance,
at very small detuning, the wave functions of these dressed states
can be approximately written as the superposition of unperturbed
conduction and valence wave function,
$\Psi=(\Psi_{+}+\Psi_{-})/\sqrt{2}=(1,0)$. These dressed states are
not the eigenstates of the helicity operator. The chiral symmetry is
broken and perfect chiral tunneling is strongly suppressed.
Numerical results are shown in Fig. 2(c) with pump intensity
$I_{\omega}=30$ MW/cm$^{2}$ and $D=300$ nm. From Fig. 2(c) we can
find that the transmission is strongly suppressed, even with lager
detuning (e.g., $\Delta_{0}=10$ meV, the transmittance is about
0.025). When detuning increases, the light-induced mixing becomes
weak [see Fig. 2(b)], the reflectance decreases, and the
transmittance increases. Fig. 2(d) shows the transmittance as a
function of pump intensity with different  barrier widths. The
strong laser field can enhance band mixing and reduce the
transmittance. From Fig. 2(d) we also see that the wide barrier can
prolong the interaction time between electrons and photons, reduce
the tunneling rate, and lower the threshold of the pump laser power.

In conclusion, we have calculated  the influence of the OSE on the
chiral tunneling in graphene by using the FDTD method. We find that
perfect tunneling can be strongly suppressed by the optically
induced band mixing, even at large detuning. These properties might be
useful in device applications, such as the fabrication of an
optically controlled field-effect transistor that has ultrafast
switching times and low power consumption.

This work was supported by the NSFC Grant Nos. 10904059,
10904097, and 11004199, the NSF from Jiangxi Province 2009GQW0017, the Open Research Fund of State Key Laboratory of Millimeter Waves No. K200901.

\end{document}